\documentclass{tlp}

\usepackage{iftex}

\ifpdf
  \usepackage{underscore}         
  \usepackage[T1]{fontenc}        
\else
  \usepackage{breakurl}           
\fi

\usepackage{amsmath}
\usepackage{graphicx}
\usepackage{listings}
\usepackage{multirow}
\usepackage{url}
\usepackage{tabularx,booktabs} 
\usepackage{bbding}            
\usepackage{changepage}        
\usepackage{xspace}

\def\lrar{\leftrightarrow}
\def\beq{\begin{equation}}
\def\eeq#1{\label{#1}\end{equation}}
\def\ba{\begin{array}}
\def\ea{\end{array}}

\def\p2f{\hbox{p2f}}

\newcommand{\selp}{{\sc selp}}

\newcommand{\cct}{{\sc ccT}}
\newcommand{\gringo}{{\sc gringo}}
\newcommand{\clingo}{{\sc clingo}}
\newcommand{\anthem}{{\sc anthem}\xspace}
\newcommand{\anthemzero}{{\sc anthem-1}}
\newcommand{\anthemone}{{\sc anthem-se}}
\newcommand{\anthemtwo}{{\sc anthem 2.0}}
\newcommand{\ap}{{\sc anthem-p2p}}
\newcommand{\vampire}{{\sc vampire}}

\let\svprovproof\proof
\let\svprovendproof\endproof
\let\proof\relax
\let\endproof\relax
\let\proof\svprovproof
\let\endproof\svprovendproof
\makeatletter
\newcommand{\printfnsymbol}[1]{%
  \textsuperscript{\@fnsymbol{#1}}%
}
\makeatother
\newcommand{\mg}{{\sc mini\nobreakdash-gringo}}
\newcommand{\ag}{{\sc abstract gringo}}

\newcommand{\dlv}{\textsc{dlv}}
\newcommand{\smodels}{\textsc{smodels}}

\newcommand{\boldp}{\mathbf{p}}

\newcommand{\boldq}{\mathbf{q}}

\newcommand{\ruleo}{\;\hbox{:-}\;}

\newcommand{\eqdef}{%
  \mathrel{\vbox{\offinterlineskip\ialign{%
    \hfil##\hfil\cr%
    $\scriptscriptstyle\mathrm{def}$\cr%
    \noalign{\kern1pt}%
    $=$\cr%
    \noalign{\kern-0.1pt}%
}}}}

\def\rar{\rightarrow}
\def\lrar{\leftrightarrow}

\def\ba{\begin{array}}
\def\ea{\end{array}}
\def\bce{\begin{center}}
\def\ece{\end{center}}

\def\beq{\begin{equation}}

\lefttitle{Fandinno et al.}

\jnlPage{1}{42}
\jnlDoiYr{20o1}
\doival{10.1017/xxxxx}

\begin{document}

\title[\anthem\ 2.0: Automated Reasoning for Answer Set Programming]{\anthem\ 2.0: Automated Reasoning for\\Answer Set Programming}

\begin{authgrp}
  \author{\sn{Jorge} \gn{Fandinno}, \sn{Zachary} \gn{Hansen}, \sn{Yuliya} \gn{Lierler}}
  \affiliation{University of Nebraska Omaha, Nebraska, USA}
  \and
  \author{\sn{Christoph} \gn{Glinzer}, \sn{Jan} \gn{Heuer}, \sn{Torsten} \gn{Schaub},\\\sn{Tobias} \gn{Stolzmann}}
  \affiliation{University of Potsdam, Potsdam, Germany}
  \and
  \author{\sn{Vladimir} \gn{Lifschitz}}
  \affiliation{University of Texas at Austin, Texas, USA}
\end{authgrp}

\maketitle

\begin{abstract}
  \anthem\ 2.0 is a tool to aid in the verification of logic programs
  written in an expressive fragment of \clingo 's input language named \mg,
  which includes arithmetic operations and simple choice rules but not aggregates.
  It can translate logic programs into formula representations in the logic of here-and-there,
  and analyze properties of logic programs such as tightness.
  Most importantly, \anthemtwo\ can support program verification by
  invoking first-order theorem provers to confirm that a program adheres to a first-order specification,
  or to establish strong and external equivalence of programs.
  This paper serves as an overview of the system's capabilities.
  We demonstrate how to use \anthemtwo\ effectively and interpret its results.
\end{abstract}

\section{Introduction}
\label{sec:intro}
We present \anthemtwo, the latest system developed as part of the ``Answer Set Programming + Theorem Proving'' (\anthem) project.
This system consolidates the findings and functionalities of previous prototypes into a
stable tool for verifying programs in the paradigm of Answer Set Programming (ASP).
The \anthemtwo\ system (in the sequel we drop 2.0 when referring to this system) can process ASP programs written in a
subset of the input language of the grounder \gringo~\citep{gehakalisc15a}.
This subset, described by~\cite{falilusc20am}, is referred to as~\mg.
\mg\ programs can contain arithmetic operations and simple choice rules, but not aggregates.
We can use \anthem\ to verify two important equivalence relations between
ASP programs: strong equivalence and external equivalence.
ASP programs $\Pi_1$ and $\Pi_2$ are called \emph{strongly equivalent} if
replacing~$\Pi_1$ by~$\Pi_2$ within any larger program does not affect its
stable models~\citep{lipeva01a}.
For example, the one-rule \emph{successor} programs
\begin{gather}
  \label{eq:successor}
  \text{\texttt{q(X+1) :- p(X).} \qquad\qquad and \qquad\qquad \texttt{q(X) :- p(X-1).}}
\end{gather}
are strongly equivalent.
The systems for verifying strong equivalence designed in the past~\citep{janoik04bm,val04,chlili05am,oetseitomwol09}
are limited to ground programs;
\anthem\ does not suffer this limitation.

\begin{figure}[h]
\begin{lstlisting}[
    caption = {A \textsc{mini-gringo} program, \texttt{primes.1.lp}.},
    label={list:primes.1},
    basicstyle=\ttfamily\normalsize,
    numbers=left,
    stepnumber=1,
    xleftmargin=3.5ex,
  ]
composite(I*J) :- I = 2..b, J = 2..b.
prime(I) :- I = a..b, not composite(I).
\end{lstlisting}%
\end{figure}%
\emph{External equivalence} is equivalence with respect to a ``user guide''---
a description of how the programs are meant to be used~\citep{fahalilite23am}.
Consider, for instance, the program\footnote{Most of the examples in this paper can also be found in the \texttt{res/examples} directory
of the \anthem\ repository; \url{https://github.com/potassco/anthem}.
The repository also contains installation instructions and a more detailed user manual.}
%
%
in Listing~\ref{list:primes.1},
which finds all primes within an interval $\{a,\dots,b\}$ with~$a>1$.
Its intended use can be described by the user guide in Listing~\ref{list:primes.ug}.
\begin{figure}[h]
\begin{lstlisting}[
    caption = {A user guide for the primes problem.},
    label={list:primes.ug},
    basicstyle=\ttfamily\normalsize,
    numbers=none,
    stepnumber=1,
    xleftmargin=3.5ex,
  ]
input: a -> integer.        output: prime/1.
input: b -> integer.        assumption: a > 1.
\end{lstlisting}%
\end{figure}%

It tells us that when we run the program, we are expected to specify appropriate values
for the placeholders {\tt a} and {\tt b}.
Furthermore, it says that the output of the program
consists of atoms in the stable model that contain \verb|prime/1|;
any other atoms of the stable model are auxiliary (``private'').

ASP programs $\Pi_1$ and $\Pi_2$ are (externally) equivalent with respect to a
user guide if, for any input that is permitted by the user guide,
they produce the same output.
For example, the program in Listing~\ref{list:primes.1} is externally equivalent to the more efficient program
\begin{verbatim}
  composite(I*J) :- I = 2..b, J = 2..b/I.
  prime(I) :- I = a..b, not composite(I).
\end{verbatim}
with respect to the user guide defined above,
even though the stable models of the two programs differ with respect to the auxiliary predicate \verb|composite/1|.

In the special case when programs do not accept inputs and have no auxiliary predicates,
external equivalence means simply that the programs have the same stable models.
We call this \emph{weak equivalence}.
If we allow auxiliary predicates and inputs, but we restrict the inputs to not contain placeholders, then external equivalence becomes a special case of \emph{relativized uniform equivalence with projection}~\citep{oettop08a}.

In addition to checking equivalence of programs, \anthem\ can verify
the adherence of an ASP program to a specification written in classical first-order logic.
For example, the formula below written in the custom language of \anthem\
captures the formal property encoded in the preceding example:
\begin{verbatim}
   forall X$g (prime(X$g) <-> a <= X$g <= b
              and not exists D$i M$i (1 < D$i < X$g and M$i*D$i = X$g)).
\end{verbatim}
This formula contains variables of two sorts:
a general variable named \verb|X| and integer variables \verb|D| and \verb|M|.
The sort of the variable is indicated by the suffix \verb|$g| or \verb|$i| for the sorts \emph{general} and \emph{integer}, respectively.
Analogously to how it checks external equivalence between programs,
\anthem\ can verify that the \verb|prime/1| predicate as defined by Listing~\ref{list:primes.1}
possesses the property encoded by the specification with respect to the user guide in Listing~\ref{list:primes.ug}.

%
%

Using \anthem\ to verify external equivalence between \mg\ programs involves
\begin{itemize}
  \item transforming rules into first-order sentences,
  \item forming program completions, 
  \item reducing the tasks described above to first-order theorem proving, and
  \item using the theorem prover \vampire~\citep{kovvor13} for proof search.
\end{itemize}

\cite{falilusc20am} describe the original system that pioneered some of the presented ideas.
We refer to this system as \anthemzero.
It was designed with the goal of verifying the adherence of a program to a specification written in first-order logic.
A related “translate and verify” system, \anthemone, was developed by~\cite{lilusc19am} for verifying strong equivalence
and extended to programs with negation by~\cite{heuer20}.
Finally, the \ap\ system~\citep{fahalilite23am} was an application built on top of the \anthemzero\ system for verifying external equivalence of programs.

Now, we present \anthemtwo, which combines and extends the functionalities of the previous
systems within a new, standalone library and command line application.
Our system helps users verify strong, external, and weak equivalence of programs,
analyze properties of programs such as tightness and regularity,
and translate programs into the syntax of many-sorted first-order logic.
The theory behind \anthem\ has been given a thorough treatment in previous publications, 
and we refer to these publications for theoretical concepts instead of redefining them here.

\section{Preliminaries: Program and ``Target'' Languages}
\label{sec:mg.language}
\label{sec:target.language}

The \ag~(AG) language~\citep{gehakalisc15a} is a theoretical representation of the input language accepted by the widely used ASP solver \clingo.
The fragments of AG studied in the context of \anthem\ are commonly referred to
as \mg\ --- this is the subset of AG for which an appropriate translation to 
the syntax of first-order logic has been widely studied. 
In this paper, we follow the definition presented by~\cite{falilusc20am} when we refer to \mg.
A \mg\ program consists of basic rules, choice rules, and constraints:
\begin{align*}
  H\ruleo B_1,\dots,B_n.\qquad\qquad\qquad
  \{H\}\ruleo B_1,\dots,B_n.\qquad\qquad\qquad
  \ruleo B_1,\dots,B_n.
\end{align*}
Here $H$ is an atom and each $B_i$ ($1 \leq i \leq n$) is either
an atom, possibly preceded by one or two negation-as-failure symbols, or
a comparison.
The fundamental task of \anthem\ consists of verifying the correctness of \mg\ programs (Section~\ref{sec:verifying}).
To achieve this goal, the system implements multiple translations between representations of these programs in different languages.
One of these languages is the TPTP format, which is a standard format for Automated Theorem Prover (ATP) systems~ \citep{sut17}.
Hence, the different verification tasks that \anthem\ can perform are based on the translation of \mg\ programs into a language of logical formulas,
running  an ATP system on a task assembled from the translated program(s),
and interpreting the results of the ATP system to provide an answer to the verification task.
%
Transforming an equivalence claim about \mg\ programs into a series of TPTP problems
is a non-trivial process and in general requires several intermediate transformations.
Thus, at the heart of \anthem\ is a logical language, which we call here simply the \emph{target language}, that supports all these intermediate transformations.
Theories written in this language may be interpreted under
the semantics of classical first\nobreakdash-order logic or
the logic of \emph{here-and-there} (HT;~\citealt{heyting30a}),
but \emph{syntactically} this is
a first-order language with variables of three sorts:
(1) a sort whose universe contains integers, symbolic constants, and special symbols \texttt{\#inf} and \texttt{\#sup},
(2) a subsort corresponding to integers, and
(3) a subsort corresponding to symbolic constants.
%

Variables ranging over these sorts are written as ``\texttt{Name\$sort}'',
where \texttt{Name} is a capitalized word and \texttt{sort} is one of the sorts defined above.
Certain abbreviations are permitted:
\begin{itemize}
  \item a general variable named \texttt{V} can be written as \texttt{V}, \texttt{V\$g}, or \texttt{V\$general};
  \item an integer variable named \texttt{X} can be written as \texttt{X\$}, \texttt{X\$i}, or \texttt{X\$integer}; and
  \item a symbol variable named \texttt{S} can be written as \texttt{S\$s} or \texttt{S\$symbol}.
\end{itemize}
%
Like \ag\ and \mg, the target language is a theoretical language ---
a formula written in this language can be printed as a series of Unicode characters in different ways.
For example, the target language formula%
\begin{gather}
\forall X ( \exists I (I = X \wedge p(I)) \rar q(X))
\label{eq:ex1}
\end{gather}
 (in which~$X$ is a general variable and~$I$ is an integer variable)
 expresses that $q$ holds for all $X$ such that $X$ is an integer for
 which $p$ holds.
This formula can be read or displayed in a custom \anthem\ syntax as follows:
\begin{verbatim}
  forall X ( exists I$ (I$ = X and p(I$)) -> q(X)).
\end{verbatim}
\anthem\ can also display target language formulas in the TPTP syntax.
The preceding formula~\eqref{eq:ex1} is formatted in TPTP as
\begin{verbatim}
  ![X: general]: ( ?[I: $int]: ( ( f__integer__(I) = X ) &
                                  p(f__integer__(I)) ) => q(X) ).
\end{verbatim}

In Section~\ref{sec:verifying}, we show how verification tasks are handled by
extending \anthem 's custom syntax for target language formulas with
meta-level declarative statements controlling proof search --- we call this the \emph{control language}.

\section{Translating Logic Programs and Formulas With \anthem}
\label{sec:translating}

\anthem\ supports (1) translations from ASP programs into logical formulas, and
(2) transformations of logical formulas within the target language.

\subsection{Translating ASP Programs Into the Target Language}
\label{sec:translating.programs}

Past versions of the \anthem\ system~\citep{falilusc20am,heuer20} relied exclusively on a translation known as $\tau^*$~\citep{lilusc19am} that transformed \mg\ programs into their formula representations.
To accommodate features such as partial arithmetic functions, $\tau^*$ produces complex formulas that are not easily understood.
However, a broad fragment of \mg\ programs does not require such complexity.
For a class of rules called \emph{regular},
a more natural translation~$\nu$ has been developed to produce formulas that are more human-readable
while maintaining equivalence to their $\tau^*$ counterparts~\citep{lifschitz21am}.
%
\anthem\ employs both the $\tau^*$ translation\footnote{
In reality, $\tau^*$ is an overloaded term referring to several iterations on the idea of translating between fragments of AG and many-sorted formulas.
The definition used here follows a slightly corrected version of the seminal
\anthemzero\ publication that can be found at \url{https://arxiv.org/abs/2008.02025}.
}
and natural translation $\nu$.
%
%
%
For example, the natural translations of successor programs~\eqref{eq:successor} are
\begin{verbatim}
  forall X$i (p(X$i) -> q(X$i + 1)), forall X$i (p(X$i - 1) -> q(X$i)).
\end{verbatim}
%
%
%


%
%
%
%

For rules that do not meet the requirements of regularity, \anthem\ falls back on~$\tau^*$
to obtain their formula representations.
%
%
Translating nonregular rules is a non-trivial task;
for example, consider the rule
\begin{verbatim}
  p(X,Y) :- X / Y > 0.
\end{verbatim}
which is not regular due to the partial function of division.
%
We can apply translation $\tau^*$ to this rule using the command
\begin{verbatim}
  anthem translate <program> --with tau-star
\end{verbatim}
which produces the output
\small
\begin{verbatim}
  forall V1 V2 X Y (V1 = X and V2 = Y and exists Z Z1 (
    exists I$i J$i Q$i R$i (I$i = J$i * Q$i + R$i and (I$i = X and J$i = Y)
      and (J$i != 0 and R$i >= 0 and R$i < J$i) and Z = Q$i)
    and Z1 = 0 and Z > Z1) -> p(V1, V2)).
\end{verbatim}
\normalsize
%
%
%
%
%
%
%
%
%
Interestingly, $\tau^*$ and $\nu$ can be safely combined
by applying $\nu$ to every regular rule and $\tau^*$ to the rules that are not regular.
Such a transformation is denoted $\mu$~\citep{fanlif23a}.
\anthem\ supports $\mu$, $\nu$, and $\tau^*$ as translation options:
\begin{verbatim}
  anthem translate <program> --with <mu | natural | tau-star>
\end{verbatim}
%
%
%

\subsection{Transformations Within the Target Language}

To support the verification of ASP programs,
\anthem\ provides two transformations that can be applied to logical formulas within the target language.
Each of these transformations provides as output a formula whose validity in classical first-order logic supports some of the verification tasks described in Section~\ref{sec:verifying}.
%
%

\subsubsection{From Here-and-there Satisfaction to Classical Satisfaction}
\label{sec:ht.classical}

Stable models of \mg\ programs can be described in terms of the logic of HT by selecting the so-called
\emph{equilibrium models}~\citep{pearce06a}.
This connection has inspired a line of research into translating programs written in ASP input languages into HT theories whose equilibrium models correspond to the answer sets of the original program.
Interestingly, for the translations concerning this paper, two programs are strongly equivalent if and only if their formula representations are equivalent in HT~\citep{lipeva01a}.
This gives us a method for verifying strong equivalence.

To reduce the task of reasoning about HT theories to reasoning over classical theories,
transformations have been proposed that ``embed'' the behavior of HT-satisfaction into classical satisfaction of  transformed formulas.
One of the first representations of this process in the ASP literature can be attributed to~\cite{petowo01am},
who describes a transformation that converts propositional HT theories into classical theories over an extended signature.
%
%
%
A generalization $\gamma$ of this translation to formulas with variables was studied by~\cite{heuer20} and by~\cite{fanlif23a}.
The $\gamma$ transformation was fundamental to the design of \anthemone, which implemented a procedure for strong equivalence checking.
For example, if $R$ is the rule
\begin{align*}
  \texttt{q(X,Y+1) :- p(X,Y).}
\end{align*}
then $\nu R$ is
\begin{gather*}
  \forall X N (p(X, N) \rar q(X, N + 1)).
\end{gather*}
where $N$ is an integer variable.
The result of applying~$\gamma$ to this formula is
\begin{gather*}
  \forall X N ((hp(X, N) \rar hq(X, N + 1)) \wedge (tp(X, N) \rar tq(X, N + 1))).
\end{gather*}
The classical models of $\gamma(\nu R)$ satisfying the additional axioms
\begin{gather*}
  \forall X Y (hp(X, Y) \rar tp(X, Y)) \qquad\text{ and }\qquad
  \forall X Y (hq(X, Y) \rar tq(X, Y))
\end{gather*}
correspond to the HT models of $\nu R$.
In terms of Kripke models, new predicates $hp/1$ and~$hq/1$ represent
satisfaction in the ``here'' world; $tp/1$ and $tq/1$ represent
satisfaction in the ``there'' world.
We call such additional axioms \emph{ordering} sentences.

  %
%

The $\gamma$ transformation plays a central role in the automated verification of strong equivalence (Section~\ref{sec:se}).

\subsubsection{Completion}
\label{sec:completion}

For the so-called \emph{tight} programs~\cite[Section 6]{falilusc20am},
Clark's Completion~\citep{clark78am} characterizes the stable models of a logic program as the models of a classical first-order theory.
%
Since its introduction, the idea of completion has been widely generalized. 
In the context of \anthem, we view completion as a transformation on top of target language theories of a certain form.
These so-called \emph{completable} theories can be transformed according to the completion procedure described by~\cite{falite24a}
into first-order theories. 
The $\tau^*$ transformation is designed to produce completable theories when applied to \mg\ programs.
For instance, we can compute the completion of the program \texttt{choice.1.lp} consisting of the rule
$\{q(X)\} \ruleo p(X)$
by passing the output of the $\tau^*$ translation to the completion operator:
%
\begin{verbatim}
  anthem translate choice.1.lp --with tau-star \
          | anthem translate --with completion
\end{verbatim}
%
This produces the output
\begin{verbatim}
  forall V1 (q(V1) <-> exists X (V1 = X and
                          exists Z (Z = X and p(Z)) and not not q(V1))).
  forall V1 (p(V1) <-> #false).
\end{verbatim}
%
Completion allows us to use first-order theorem provers to verify the external equivalence
of programs meeting certain restrictions, such as tightness (Section~\ref{sec:ee}).
We can confirm that our program is tight with the command
\begin{verbatim}
    anthem analyze choice.1.lp --property tightness
\end{verbatim}

\section{Verifying Logic Programs With \anthem}
\label{sec:verifying}

The fundamental task of \anthem\ consists of formally verifying properties of \mg\ programs.
This section describes the three verification tasks that \anthem\ can perform: \emph{strong equivalence}, \emph{external equivalence}, and \emph{specification adherence}.

\subsection{Strong Equivalence}
\label{sec:se}

The strong equivalence of two programs $\Pi_1$ and $\Pi_2$ can be established by
deriving the equivalence $\gamma(\tau^*\Pi_1) \lrar \gamma(\tau^*\Pi_2)$
(or, equivalently, $\gamma(\mu\Pi_1) \lrar \gamma(\mu\Pi_2)$)
from the associated ordering sentences.
%
\anthem\ accomplishes this by constructing a series of subproblems for an ATP system to solve.
%
\begin{figure}[htb]
\begin{lstlisting}[
    caption = {The program \texttt{transitive.1.lp}.},
    label={list:transitive.1},
    basicstyle=\ttfamily\normalsize,
    numbers=left,
    stepnumber=1,
    xleftmargin=3.5ex,
]
{q(X,Y)} :- p(X), p(Y).
 q(X,Z)  :- q(X,Y), q(Y,Z), p(X), p(Y), p(Z).
\end{lstlisting}
\end{figure}
To illustrate this process, consider the encoding of the property\\``q is transitive on the domain of p''
given in Listing~\ref{list:transitive.1}~\citep{halipeva17am},
and a refactoring of this program (\texttt{transitive.2.lp}) that replaces the second rule with the constraint
\begin{align}
    \label{eq:transitive.2}
    \texttt{:- q(X,Y), q(Y,Z), not q(X,Z), p(X), p(Y), p(Z).}
\end{align}

We can verify their strong equivalence by invoking \anthem\
with an instruction to use~$\mu$ for obtaining the HT formula representation of these two programs:
\begin{verbatim}
    anthem verify --equivalence strong transitive.{1.lp,2.lp} \
        --formula-representation mu
\end{verbatim}
%
Let us denote Rule 1 from Listing~\ref{list:transitive.1} as~$F$ (this rule is common to both programs).
Let~$G_1$ denote Rule 2 from Listing~\ref{list:transitive.1}, and~$G_2$ denote the constraint~\eqref{eq:transitive.2}. 
Furthermore, let $\mathcal{A}$ denote ordering sentences
\begin{align*}
    \forall X (hp(X) \rar tp(X)) \qquad \text{and} \qquad\forall X_1 X_2 (hq(X_1, X_2) \rar tq(X_1, X_2)).
\end{align*}
The strong equivalence of these programs can be established by showing that
$$\mathcal{A} \rar ( \gamma(\mu(\{F, G_1\})) \lrar \gamma(\mu(\{F, G_2\})) )$$
is satisfied, in the sense of classical first\nobreakdash-order logic,
by all standard\footnote{Standard interpretations give a standard treatment to operations like integer addition.} interpretations~\citep[Theorem~2]{fanlif23a}.
\anthem\ decomposes this task into four subproblems, each of which consists of verifying that one of the following implications is satisfied by all standard interpretations:
\begin{align*}
    (\mathcal{A} \wedge \gamma(\mu F) \wedge \gamma(\mu G_1)) &\rar \gamma(\mu F), \qquad
    (\mathcal{A} \wedge \gamma(\mu F) \wedge \gamma(\mu G_1)) \rar \gamma(\mu G_2),\\
    (\mathcal{A} \wedge \gamma(\mu F) \wedge \gamma(\mu G_2)) &\rar \gamma(\mu F), \qquad
    (\mathcal{A} \wedge \gamma(\mu F) \wedge \gamma(\mu G_2)) \rar \gamma(\mu G_1).
\end{align*}
\anthem\ checks that these implications (two of which are trivial) are entailed by a set of axioms describing standard interpretations using a classical ATP system, currently \vampire, and answers that the two programs are strongly equivalent
if it finds that all four implications are entailed by the axioms.

\subsection{External Equivalence}
\label{sec:ee}

Strong equivalence can sometimes be \emph{too} strong of a condition when we are comparing the behavior of two programs.
Often, we are only interested in confirming that two programs have the same output when paired with the same input.
This type of equivalence is called \emph{external equivalence}~\citep{fahalilite23am}.
External equivalence is defined with respect to a \emph{user guide}, which defines a class of
acceptable inputs to the programs, and specifies which predicates encode their output.

A user guide contains statements of three types:
input declarations,
output declarations,
and assumptions about the inputs.
An input declaration describes an input predicate or a \emph{placeholder}
(a symbolic constant that is given a value by the program's input).
%
%
%
%
Note that \anthem\ does not require a concrete valuation of placeholders ---
for instance, Listing~\ref{list:primes.ug} only specifies that \verb|a| is an integer greater than 1.
This user guide also specifies, via the declaration of \texttt{prime/1} as the only output predicate,
that only atoms in the stable model that contain \texttt{prime/1} are considered when checking for external equivalence.
\begin{figure}[h]
\begin{lstlisting}[
      caption = {A refactored primes program, \texttt{primes.3.lp}.},
      label={list:primes.3},
      basicstyle=\ttfamily\normalsize,
      numbers=left,
      stepnumber=1,
      xleftmargin=3.5ex,
    ]
sqrtb(M) :- M = 1..b, M*M <= b, (M+1)*(M+1) > b.
composite(I*J) :- sqrtb(M), I = 2..M, J = 2..b.
prime(I) :- I = a..b, not composite(I).
\end{lstlisting}
\end{figure}
We can verify that the program in Listing~\ref{list:primes.1} is externally equivalent
to the more efficient program in Listing~\ref{list:primes.3} for \emph{any} pair of integers $(a,b)$ such that $a > 1$.
If \texttt{primes.ug} is the file containing the user guide in Listing~\ref{list:primes.ug},
\anthem\ can automatically prove the aforementioned claim.
%

To achieve this, \anthem\ first builds the completion of the $\tau^*$ formula representation of each of the programs (Sections~\ref{sec:translating.programs} and~\ref{sec:completion})
and then produces a series of subtasks to verify using \vampire.
These subtasks are described by~\cite{fahalilite23am} and a detailed example is given in Section~\ref{sec:outlines}.
Note that verifying the preceding example is difficult for the current version of \vampire\ without some human help (see Section~\ref{sec:practice}).
It is also worth mentioning that currently, \anthem\ can only automatically verify external equivalence for programs that are tight and lack private recursion~\cite[Section 6]{fahalilite23am}.
Note that strong equivalence verification does not suffer this restriction,
and, in certain cases, the tightness limitation can be lifted (see Section~\ref{sec:practice}).




\subsection{Specification Adherence}
\label{sec:spec.adherence}

Thus far, we have discussed how \anthem\ can be used to establish that certain types
of equivalences hold between two ASP programs.
In this section, we show how \anthem\ can verify the adherence of an ASP program to a specification
written in classical first-order logic.
%
This workflow can be viewed as a special type of external equivalence checking, where instead
of encoding our specification of desired behavior as an ASP program, we encode it directly in \anthem 's control language.
\begin{figure}[h]
\begin{lstlisting}[
      caption = {An encoding (\texttt{cover.lp}) solving the exact cover problem.},
      label={list:cover},
      basicstyle=\ttfamily\normalsize,
      numbers=left,
      stepnumber=1,
      xleftmargin=3.5ex,
    ]
{in_cover(1..n)}.
:- I != J, in_cover(I), in_cover(J), s(X,I), s(X,J).
covered(X) :- in_cover(I), s(X,I).
:- s(X,I), not covered(X).
\end{lstlisting}
\end{figure}
For example, we can validate that the program given in Listing~\ref{list:cover}
correctly solves the exact cover\footnote{
    An exact cover of a collection $S$ of sets is a subcollection $S'$ of $S$ such that each element of the union of all sets in $S$ belongs to exactly one set in $S'$.
}
problem by directly encoding the properties our program should possess as follows.
%

First, a solution to this problem is a collection of set identifiers in the range $[1,n]$:
\begin{verbatim}
    spec: forall Y (in_cover(Y) ->
            exists I$ (Y = I$ and I$ >= 1 and I$ <= n$i)).
\end{verbatim}
Second, each element that occurs in the union of all sets must occur in the cover:
\begin{verbatim}
    spec: forall X (exists Y s(X, Y) ->
                        exists Y (s(X, Y) and in_cover(Y))).
\end{verbatim}
Finally, sets selected to be part of the cover cannot overlap:
\begin{verbatim}
    spec: forall Y Z (exists X (s(X, Y) and s(X, Z))
                        and in_cover(Y) and in_cover(Z) -> Y = Z).
\end{verbatim}
Keep in mind that we need to restrict our input to sets identified by the range $[1,n]$:
\begin{verbatim}
    assumption: forall Y (exists X s(X, Y) ->
                    exists I$ (Y = I$ and I$ >= 1 and I$ <= n$i)).
\end{verbatim}
We can verify Listing~\ref{list:cover} against this specification with the following user guide:
\begin{verbatim}
    input: n -> integer.             input: s/2.
    output: in_cover/1.              assumption: n >= 0.
\end{verbatim}
using the command
\begin{verbatim}
    anthem verify --equivalence external cover.{lp,spec,ug}
\end{verbatim}
This workflow tells us that \texttt{cover.lp} has the properties encoded by the specs,
and that our specification \texttt{cover.spec} completely defines the external behavior of the program in \texttt{cover.lp} with respect to this user guide.
If we remove one of the specs from \texttt{cover.spec}, then the preceding command fails, since the specification is no longer a complete description of the program's behavior.
However, the following command succeeds.
\begin{verbatim}
    anthem verify --equivalence external cover.{lp,spec,ug} \
                  --direction backward
\end{verbatim}
This tells us that the program embodies the remaining properties of the specification.
%


More broadly, for any verification task, \anthem\ attempts to validate some form of equivalence.
Thus, we can run the \texttt{forward} and \texttt{backward} directions of this equivalence proof separately if desired.
In the case of external equivalence, we can frame our task as a proof of the equivalence between a program and a specification;
our specification can be written as a logic program,
or as a collection of assumptions about the input and formulas annotated with the \emph{spec} role.

\section{\anthem in Practice}
\label{sec:practice}

One of the goals of this paper 
is to provide practical advice for using \anthem,
and demystify the interpretation of the results produced.
Chances are your experience with \anthem\ will fall into one of three categories.

\subsection{The Good}

This is the most straightforward case: \anthem\ returns a message like
\begin{verbatim}
  > Success! Anthem found a proof of the theorem.
\end{verbatim}
For instance, invoking the command from Section~\ref{sec:se} 
produces an output summarizing the two subproblems passed to \vampire\ and their respective success statuses.
%
Here, all subproblems in both the forward and backward directions of the equivalence proof ended with a ``Status: Theorem'' message.
This means that the subproblem conjecture was successfully derived from the subproblem axioms.
Since every subproblem was verified, the programs \texttt{transitive.1.lp} and \texttt{transitive.2.lp}
are strongly equivalent.

\subsection{The Bad}
There are certain types of problems that \anthem\ is not equipped to address.
The most common such problem is the task of validating the external behavior of a non-tight program, 
%
%
which the current incarnation of \anthem\ refuses to attempt.
One notable exception is the case of \emph{local tightness}~\citep{falite24a}.
Many non-tight programs are still locally tight, such as programs encoding planning problems in which the law of inertia introduces positive recursion.
While tightness is a syntactic property of programs that is easily checked by \anthem,
an effective procedure for checking local tightness has not been developed.
If, however, a user has manually proven the local tightness of their program(s),
\anthem's tightness check can be bypassed, and verification safely completed, by adding the
\verb|--bypass-tightness| flag.


\subsection{The Ugly}
\label{sec:outlines}

When \anthem\ fails to verify a program, it is \emph{not} a proof that the programs are not (strongly, externally) equivalent.
It very well may be that the proof exists, but \anthem\ needs some help finding it.
What can be done in these cases?

\textbf{Option 1:}
Increase resource allocation.
\anthem\ lets you increase the timeout \verb|-t| for the backend ATP for each subproblem.
You can also parallelize search by
increasing the number of cores used by the ATP with the \verb|-m| flag.

\textbf{Option 2:}
Explore missing or malformed assumptions.
If \anthem\ is hung up on a particular subproblem, consider the axiom set and conjecture.
Is it clear that the conjecture follows from the axioms?
Sometimes seemingly self-evident assumptions are missing --- for a detailed example, see the
``orphan'' example by~\cite{fahalilite23am}.
%

\textbf{Option 3:}
Write a proof outline.
If \vampire\ is unable to validate a subproblem in a reasonable amount of time (as is often the case for nontrivial tasks),
the next step is to find a lemma that can be
derived by \vampire\ from the axioms without help and that is likely to facilitate
achieving the goal when added to the list of axioms.
Thus, for external equivalence tasks, \anthem\ lets users supply a \emph{proof
outline} 
consisting of annotated formulas that can play three roles: definitions, lemmas, and inductive lemmas.
%
%
These are target language formulas augmented with instructions for how they should be used
within a verification task.
Their general form is
\begin{verbatim}
  role(direction)[name]: formula.
\end{verbatim}
where \texttt{role} and \texttt{formula} are required,
and \texttt{direction} indicates which direction of the equivalence proof the formula is used within.
Such outlines can often be very effective.

\paragraph{Definitions}%
are assumed to define the extent of a
new predicate introduced for convenience within a proof outline.
They have the form
\begin{verbatim}
  definition: forall X ( p(X) <-> F(X) ).
\end{verbatim}
where \verb|X| is a tuple of variables,
\verb|p| is a fresh predicate symbol, and
\verb|F| is a formula with free variables \verb|X|.
%
%
A sequence of definitions is valid if any defined predicate $p$~used within each~$F$ is defined previously in the sequence.
For example, the annotated formula ($D$)
\begin{verbatim}
    definition[D]: forall I$ N$ (sqrt(I$,N$) <->
                  I$ >= 0 and I$*I$ <= N$ < (I$+1)*(I$+1)).
\end{verbatim}
defines the integer square root ($I$) of an integer $N$.
Definitions can be used in a proof similarly to assumptions as their purpose is to make writing lemmas easier.

\paragraph{Lemmas.}
\anthem\ interprets proof outlines in an intuitive way: as a series of intermediate claims to
be checked en route to a final claim.
Previously established results in this sequence are used as assumptions when checking the next claim.
%
%
When \anthem\ is provided with a proof outline as part of an external equivalence verification task,
it attempts to sequentially verify every (inductive) lemma.
If successful, it treats the resulting set of formulas as assumptions during the verification task.
For example, a proof outline consisting of $D$ and the following two lemmas ($L_1$ and $L_2$)
is interpreted as an instruction to derive $L_1$ from $D$ and $L_2$ from $\{D, L_1\}$.
\begin{verbatim}
  lemma[L1]: sqrt(I$,N$) and (I$+1)*(I$+1) <= N$+1 -> sqrt(I$+1,N$+1).
  inductive-lemma[L2]: N$ >= 0 -> exists I$ sqrt(I$,N$).
\end{verbatim}

\paragraph{Inductive lemmas}%
have the general form
\begin{verbatim}
  inductive-lemma: forall X N$ ( N$ >= n -> F(X,N$) ).
\end{verbatim}
where \verb|n| is an integer, \verb|X| is a tuple of variables, \verb|N| is an integer variable and \verb|F| is a target language formula.
Within a proof outline, an inductive lemma is interpreted as an instruction to prove two conjectures:
\begin{gather*}
   \forall X  F(X,n) \qquad \text{ and } \qquad \forall X N ( N \geq n \wedge F(X,N) \rar F(X,N+1) ).
\end{gather*}
If both the first (the base case) and the second (the inductive step) conjectures are proven,
then the original formula
\begin{gather*}
  \forall X N ( N \geq n \rar F(X,N) )
\end{gather*}
is treated as an axiom in the remaining proof steps.
For example, $L_2$ is interpreted as an instruction to verify
\begin{gather*}
  \exists I sqrt(I,0) \qquad \text{ and } \qquad \forall N (N \geq 0 \wedge \exists I sqrt(I,N) \rar \exists J sqrt(J,N+1)).
\end{gather*}

We can verify the external equivalence of Listings~\ref{list:primes.1} and~\ref{list:primes.3}
with respect to the user guide in Listing~\ref{list:primes.ug} with the proof outline \verb|primes.po|,
which extends $\{D, L_1, L_2\}$ with the following lemmas, $L_3$ and $L_4$:
\begin{verbatim}
  lemma[L3]: b >= 1 -> (sqrtb(I$) <-> sqrt(I$,b)).
  lemma[L4]: I$ >= 0 and J$ >= 0 and I$*J$ <= b and sqrtb(N$)
                                    -> I$ <= N$ or J$ <= N$.
\end{verbatim}

Let us denote the assumption \verb|a > 0| from Listing~\ref{list:primes.ug} as $A$.
Additionally,
let us denote the completion of Rule~1 from Listing~\ref{list:primes.1} as~$F$,
the completion of Rule~2 from Listing~\ref{list:primes.1} as $F'$,
the completion of rules 1 and 2 from Listing~\ref{list:primes.3} as $G$,
and the completion of Rule~3 from Listing~\ref{list:primes.3} as $G'$.
%
When the following command is invoked
\begin{verbatim}
  anthem verify --equivalence external primes.{1.lp,3.lp,ug,po}
\end{verbatim}
\anthem\ verifies the proof outline as described above,
then verifies the following subproblems
in which the completed definitions of private predicates are treated as assumptions from which
to derive the equivalence of the public predicates' definitions:
%
\begin{align*}
  &\left( A \wedge L_1 \wedge L_2 \wedge L_3 \wedge L_4 \wedge F(\boldp) \wedge G(\boldq) \wedge  F'(\boldp) \right) \rar G'(\boldq),\\
  &\left( A \wedge L_1 \wedge L_2 \wedge L_3 \wedge L_4 \wedge F(\boldp) \wedge G(\boldq) \wedge  G'(\boldq) \right) \rar F'(\boldp).
\end{align*}
Here, $\boldp$ is a list of fresh predicate symbols, $\{composite/1\}$, to replace the private predicates occurring in \verb|primes.1.lp|.
Similarly, $\boldq$ replaces the private predicates occurring in \verb|primes.3.lp| with predicate symbols that won't conflict with $\boldp$: $\{sqrtb/1, composite\_p/1\}$.




\paragraph{A Challenging Example}
\cite{falite24a} provide two alternative solutions (Sections~1 and 5)
to the frame problem, using different approaches to encoding the law of inertia.
While these programs are not tight, they are locally tight, allowing us to safely bypass the tightness check.
If we provide \anthem\ with
the following proof outline
\begin{verbatim}
  inductive-lemma(backward): N$ >= 0 -> (in(X,Y,N$) -> person(X) )
\end{verbatim}
%
%
we can verify the external equivalence of the two frame programs.
%
This example is remarkable due to the complexity of the programs
(which use powerful modeling features to encode a realistic problem from the ASP literature),
and the fact that we can prove external equivalence for arbitrary time horizons $h \geq 0$ using a single inductive lemma.

\section{Experimental Analysis}


This section compares the capabilities (Table~\ref{table:lit.verification.tools})
and performance (Table~\ref{table:benchmarking}) of \anthemtwo\ against related systems,
particularly its predecessors.
As mentioned in the Introduction, tools from the \anthem\ family are oriented towards a different class of logic programs
than \selp, \textsc{dlpeq}, and \cct, which were designed for propositional \smodels\ or \dlv\ programs with disjunctive heads.
As shown in Table~\ref{table:lit.verification.tools}, they do not support variables, double negation, or arithmetic.
Hence, there cannot be a sensible runtime comparison between these systems and \anthemtwo.
Furthermore, they employ a different type of backend inference engine --
the common theme of these systems is to convert a verification task into a satisfiability problem for some form of solver.
\selp, \textsc{dlpeq}, and \cct\ use SAT, ASP, and Quantified Boolean Formula (QBF) solvers, respectively.
\anthem\ systems, on the other hand, use an automated theorem prover (ATP) as a backend.

In addition to supporting several useful language features as shown in Table~\ref{table:lit.verification.tools},
\anthemtwo\ also permits users to verify multiple types of equivalence.
For example, it supports specification adherence verification in the style of \anthemzero,
implemented as a special case of external equivalence.
This type of verification is unique -- all other types in the table refer to a form of equivalence between logic programs.
Interestingly, \ap\ and \anthemtwo, by virtue of their ability to verify external equivalence of logic programs,
also support weak equivalence. 
Weak (or answer set) equivalence between programs $\Pi_1$ and $\Pi_2$ can be verified
by comparing them under a user guide without placeholders, assumptions or input declarations,
that contains every predicate in $\Pi_1 \cup \Pi_2$ as an output predicate.
%
%
Note that \cct\ supports a form of relativized uniform equivalence with projection
similar to external equivalence when placeholders are forbidden,
in addition to a generalized form of strong equivalence relativized to a propositional signature.

\small
\begin{table}[]
    \begin{adjustwidth}{-4cm}{-4cm}
    \small
    \centering
    \setlength\tabcolsep{3.5pt}
    \begin{tabular}{|l|c|c|c|c|c|c|c|c|c|l|}
        \hline
        Tool & \multicolumn{4}{c|}{Language Features} & \multicolumn{4}{c|}{Equivalence} & \multicolumn{1}{c|}{Backend} \\

          & Vars. & Disj. & Dbl. Neg. & Arith. & Strong & External & Weak & Spec & \\
        \hline
        \selp &  &  \Checkmark &  &  & \Checkmark & & & &  SAT \\
        \textsc{dlpeq} &  &  \Checkmark &  &  & \Checkmark & & \Checkmark &   & ASP \\
        \cct &  &  \Checkmark &  &  & \Checkmark & & \Checkmark  &  &  QBF \\
        \anthemone & \Checkmark &  & \Checkmark & \Checkmark & \Checkmark & &  & & ATP \\
        \anthemzero & \Checkmark &   & \Checkmark & \Checkmark &   & &  & \Checkmark  & ATP \\
        \ap & \Checkmark &   & \Checkmark & \Checkmark &   & \Checkmark & \Checkmark &  & ATP \\
        \anthemtwo & \Checkmark &   & \Checkmark & \Checkmark &  \Checkmark & \Checkmark & \Checkmark  & \Checkmark  & ATP \\
        \hline
    \end{tabular}
    \caption{Tools supporting proof-based verification of ASP programs.}
    \label{table:lit.verification.tools}
    \end{adjustwidth}
\end{table}
\normalsize

Recall that \anthemtwo\ was an effort to integrate, stabilize, and extend the capabilities
of previous prototypes developed for one type of equivalence only.
As such, there is at most one competitor to \anthemtwo\ for each problem in Table~\ref{table:benchmarking}.
The problems considered are a subset of the repository's \textbf{res/examples} folder,
excluding only trivial parsing tests.
%
%
Times are given in seconds, the $m$ parameter represents the number of cores allocated to \vampire,
and the ``PO'' column indicates if a proof outline was provided to aid in solving the problem within the 5 minute time limit.
All experiments were conducted on Ubuntu 24.04.2 LTS, 13th Gen Intel(R) Core(TM) i7-1370P, 32 GB RAM.

These comparisons demonstrate that \anthemtwo\ is considerably more powerful than its predecessors when applied to non-trivial problems,
both in terms of capabilities and performance.
%
%
In particular, the new features of proof outlines (definitions and inductive lemmas)
have enabled us to verify problems that previous systems could not address in a reasonable amount of time.
Consider the challenging Frame problem discussed at the end of Section~5.
While even \anthemtwo\ was initially unable to address this problem within the 5 minute timeout,
the use of a proof outline with the new inductive lemma feature brought the runtime down to 7.88 seconds.
Indeed, Division, Frame, and Primes (2v3) all rely on proof outlines with features specific to \anthemtwo.

For certain trivial problems (Bounds, Successor, Transitive), \anthemone\ outperforms \anthemtwo,
although the differences for the Transitive problem are negligible.
%
For the other two problems, this is likely due to a shortcut employed by \anthemone\ when it encounters \emph{positive programs}.
A positive rule is a basic rule or constraint whose body does not contain negation. 
For simple programs of this nature, the $\gamma$ transformation can be omitted when checking strong equivalence~\citep[Proposition 6]{lilusc19am}.

%
%
%
%

\begin{table}
    \begin{adjustwidth}{-4cm}{-4cm}
    \small
    \centering
    \setlength\tabcolsep{3.5pt}
    \begin{tabular}{|l|l|c|l|c|c|c|c|c|c|}
        \hline
        Problem         & Equivalence & PO & Competitor     & \multicolumn{3}{c|}{Competitor Runtime} &  \multicolumn{3}{c|}{\anthemtwo\ Runtime}\\
                        &            &        &                &  m=2  &  m=4 & m=8                  &  m=2  &  m=4 & m=8   \\
        \hline
        Coloring        & External  &            & \anthemzero &  0.15 & 0.21 & 0.20                & 0.13  & \textbf{0.12} & \textbf{0.12}  \\
        Cover (1vs)     & External  &            & \anthemzero &  0.40  & 0.22 & 0.15               & 0.11  & \textbf{0.10}  & \textbf{0.10}   \\
        Cover (1v2)     & External  &            & \ap         &  0.28 & 0.25 & 0.32               & 0.13  & \textbf{0.11} & 0.13  \\
        Division        & External  &            & \anthemzero & \multicolumn{3}{c|}{Timeout}      & \multicolumn{3}{c|}{Timeout} \\
        Division        & External  & \Checkmark &    N/A      & \multicolumn{3}{c|}{-}            & 6.20   & 3.21 & \textbf{1.68} \\
        Frame           & External  &            & \ap         & \multicolumn{3}{c|}{Timeout}      & \multicolumn{3}{c|}{Timeout} \\
        Frame           & External  & \Checkmark &   N/A       & \multicolumn{3}{c|}{-}            & 36.69 & 17.97 & \textbf{7.88} \\
        Primes (1v2)    & External  &            & \ap         &  5.57 & 7.21 & 2.94               & 2.56  & 1.14 & \textbf{0.54}  \\
        Primes (2v3)    & External  &            & \ap         & \multicolumn{3}{c|}{Timeout}      & \multicolumn{3}{c|}{Timeout} \\
        Primes (2v3)    & External  & \Checkmark &    N/A      & \multicolumn{3}{c|}{-}            & 95.57 & 48.14 & \textbf{29.47} \\
        Primes (2vs)    & External  &            & \anthemzero & \multicolumn{3}{c|}{Timeout}      & 6.44  & 3.14 & \textbf{1.77} \\
        Bounds          & Strong    &            & \anthemone  & 0.25 & 0.15 & \textbf{0.07}       & 1.80   & 0.99 & 0.48 \\
        Choice          & Strong    &            & \anthemone  & \multicolumn{3}{c|}{Timeout}      & \textbf{0.04}  & 0.09 & 0.07 \\
        Squares         & Strong    &            & \anthemone  & 17.37 & 141.43 & 69.45            & 5.71  & 2.53 & \textbf{1.20} \\
        Successor       & Strong    &            & \anthemone  & 0.86 & 0.45 & \textbf{0.25}       & 1.26  & 0.64 & 0.40  \\
        Transitive      & Strong    &            & \anthemone  & 0.06 & \textbf{0.05} & 0.06       & 0.07  & 0.11 & \textbf{0.05} \\
        \hline
    \end{tabular}
    \caption{Comparing \anthemtwo\ against its predecessors.}
    \label{table:benchmarking}
    \end{adjustwidth}
\end{table}


\section{Conclusions and Future Work}

This paper provides an overview of the \anthemtwo\ system.
We strongly encourage interested readers to also visit the \anthem\ repository,
where a user manual and set of examples with expected outputs may be found.
%
%
%
The systems developed in the \anthem\ project are unique within the landscape of (proof-based) ASP verification tools
due to their support for programs with variables and integer arithmetic.
\anthemtwo\ subsumes previous \anthem\ systems and provides powerful new features such as natural translation for strong equivalence tasks and enhanced proof outlines for external equivalence problems.
Proof outlines, for instance, make previously unsolved problems such as the external equivalence of
Listings~\ref{list:primes.1} and~\ref{list:primes.3} tractable.
%
%
Additionally, the (optional) decoupling of the translation and verification steps
may prove to be a useful tool for studying the relationship between ASP and other logical formalisms such as classical first-order logic.
%
%

This tool offers a stable foundation for future innovation.
A number of improvements are already planned or in progress, including
improved algorithms for simplifying formulas,
proof outlines for strong equivalence tasks,
employing alternative backend ATP systems,
integration of natural translation into external equivalence verification, 
bypassing tightness restrictions with ordered completion and/or tightening,
extending the supported language with conditional literals, 
supporting counting
and/or unrestricted
aggregates,
and revising the definition of integer division for consistency with \clingo~\cite[Footnote 3]{fanlif23}.
The long-term goal is to support the complete \ag\ language within \anthem.

\section*{Acknowledgements}
We are thankful to the anonymous reviewers for their valuable feedback.
This research is partially supported by NSF CAREER award 2338635
and the research development program at the University of Nebraska Omaha.
Any opinions, findings, and conclusions or recommendations expressed in this material are those of the authors and do not necessarily reflect the views of the National Science Foundation.

\bibliographystyle{plainnat} 
\bibliography{misc,krr,procs}

\end{document}